\documentclass{an}
\usepackage{ae}
\usepackage{graphicx}
\usepackage{times}

 \usepackage[latin1,ansinew]{inputenc}
 \usepackage[T1]{fontenc}
 \usepackage{placeins}

\newcommand{\dif}{\ensuremath{\mbox{d}}}
\renewcommand{\imath}{\rm i}

\begin{document}

\sloppy


\title{Estimating the Solar Meridional Circulation by Normal Mode Decomposition}

\author{Lars Krieger\inst{1}%
\and  Markus Roth\inst{2,1}
\and Oskar von der L\"uhe\inst{1}
}
\titlerunning{Estimation of the solar meridional circulation}
\authorrunning{Lars Krieger \& Markus Roth \& Oskar von der Lühe}
\institute{
Kiepenheuer-Institut für Sonnenphysik, Schöneckstraße 6,
79104 Freiburg, Germany
\and
Max-Planck-Institut für Sonnensystemforschung, Max-Planck-Straße 2, 37191
Katlenburg-Lindau, Germany
}
\received{30 May 2005}
\accepted{11 Nov 2005}
\publonline{later}


\abstract{ The objective of this article is to use Fourier-Hankel
decomposition as suggested earlier by~\cite{fanbraun} to estimate the
integrated horizontal meridional flow velocity as a function of mode
penetration depth, and to find ways of potentially improve this technique.
We use a time series of 43200 (30 days) consecutive full-disk Dopplergrams
obtained by the MDI (Michelson Doppler Imager) instrument aboard the SOHO
(Solar Heliospheric Observatory) spacecraft in April 1999. We find
averaged meridional flow estimates of 15~m/s for modes with a penetration depth in the
upper 20~Mm of the solar convection zone. This reproduces the results of
the earlier investigations. Moreover we conclude that this method has the
potential to become a new technique to measure the meridional circulation in
the deep convection zone, if some improvements will be applied.}

\maketitle

\section{Introduction}

On the solar surface a meridional flow of 10--20~m/s can be
measured. This flow is directed away from the equator towards the
poles~(\cite{hatha,komm,latu,snod}). Because
of mass conservation a poleward meridional flow in the outer part
of the convection zone requires an equatorward return flow in deeper
layers.  This meridional circulation might play an
important role in the dynamo process which causes the solar magnetic
cycle~(\cite{choud,dik1,dik2,bb}). It has been
the topic of various studies (starting with~\cite{giles97}) to
infer the depth-dependent profile of this meridional circulation with
various techniques from helioseismology.

The motivation for this work is to reproduce a method suggested
by~\cite{fanbraun} for inferring the speed of the averaged solar meridional flow.
The method
is based on splitting up the oscillation signal observed in a time series of
Dopplergrams into two wave-fields of anti-parallel travelling waves by a
Fourier-Hankel decomposition~(\cite{braun87}). Originally an average meridional flow
in the order of 10~m/s was found in the upper part of the convection zone. We
are able to reproduce the results on an independent data set and with a newly
created data analysis pipeline. The reproduction of the method provides a
first step in measuring the meridional circulation. We discuss potential
improvements of this method to obtain more accurate results and to
possibly reach to greater depths in future studies.

\section{Methods and data}
\subsection{Normal mode decomposition}

\cite{braun87} applied Fourier-Hankel spectral analysis to
decompose locally the solar oscillation signal into wave fields travelling in
opposite directions. An overview to the technique is given
in~\cite{gizon05}. This decomposition was used for
investigating the influence of sunspots on the helioseismic p-modes. It
allowed detecting phase-shifts and damping of p-modes by
sunspots~(\cite{braun88, bogdan, braun95}).

A further application allowed studying the effect of the meridional flow on
the solar oscillations~(\cite{fanbraun}). However, the Fourier-Hankel
decomposition  was originally designed for studying signals on a small
section of a spherical surface by approximating the surface by a planar
geometry. Therefore it is generally not suited for studying global flows on a
sphere. Nevertheless, we reproduce the data analysis in the following. But we
note, this method can only be seen as an initial approximative step to obtain
rough estimates of average effects. It might allow to conclude on what to
expect from a more proper analysis in future.

The velocity amplitude signal of the solar eigenoscillations $\Psi(\theta,\varphi,t)$,
consisting of standing p-modes, can be locally separated into two fields of travelling
waves on the sphere. This local decomposition is carried out within an
annular region around a central point of interest. This point is the
piercing point of the axis of the used spherical coordinate system, where $\theta$ is the
latitude, and $\varphi$ is the longitude. In the original
concept~(\cite{braun87})
this point has been the center of a sunspot or an active region. In the following we
investigate a subsection of an annular region around the poles of the sun, i.e. the
polar axis of the coordinate system is identical with the rotation axis of the
sun. The coordinates used therefore coincide with the standard spherical coordinate system.

According to~\cite{fanbraun} the oscillation velocity signal $\Psi$
is expanded as
 \begin{eqnarray}
\Psi(\theta,\varphi,t)&=&\sum \limits_{lm\nu} e^{\imath(m\varphi + 2\pi\nu t)}\\
&&\times \left[A_{lm\nu}\Theta_{l}^{m}(\cos \theta) + B_{lm\nu}(\Theta_{l}^{m})^*(\cos \theta) \right] \
,
\nonumber
\end{eqnarray}
with
\begin{equation}
\Theta_{l}^{m}(\cos \theta):=N_l^m\left[P_l^m(\cos \theta) + \frac{2 \imath}{\pi}Q_l^m(\cos \theta)  \right] \
,
\end{equation}
where the asterisk denotes complex conjugation, $P_l^m$ and $Q_l^m$ are Legendre functions of the first and second kind, and $N_l^m$ is a
normalisation constant.
The complex quantities $A_{lm\nu}$ and $B_{lm\nu}$ are the amplitudes of the inward and outward going waves respectively, $n$ is the radial order, $l$ is the
harmonic degree, $m$ is the azimuthal order, $t$ is time and $\nu$ is the temporal frequency.
We refer the reader to~\cite{fanbraun} and~\cite{braun88} for more details
on these decomposition.

Following~\cite{fanbraun} Hankel functions are used as approximations to $(P_l^m\pm 2\imath /\pi Q_l^m)$
\begin{eqnarray}
H_m^{(1,2)}(L\theta)&\approx& (-1)^m
\frac{(l-m)!}{(l+m)!}\nonumber\\
&\times&  \left[P_l^m(\cos\theta)
\pm \frac{2\imath}{\pi}Q_l^m(\cos\theta)\right]\ ,
\end{eqnarray}
with $L=\sqrt{l(l+1)}$.
This approximation is valid in the limit
$l\gg m$, which is fulfilled in the current
investigation~(\cite{braun95}).  A decomposition into wavefields by
applying a method built on Legendre functions, rather than
Hankel functions, would be more accurate for the spherical case.
But the application of Hankel functions provides an estimation which
is much faster to calculate.

According to~\cite{gizon05} in some cases a discrete set of Hankel functions
can be selected which guarantees orthogonality. Explicitly an integration
over the entire range of $\varphi$ would provide such an orthogonal set of
functions in the present case. In the further data analysis, however, the
integral is carried out only over a limited range in $\varphi$. This then
introduces possibly leakage effects between modes of different azimuthal
order $m$. Therefore only the best possible approximation to an orthogonal
set can be actually selected.

The wave amplitudes $A_{lm\nu}$ and $B_{lm\nu}$ are extracted from the wave
field $\Psi(\theta,\varphi,t)$ according to~\cite{braun92} by
\begin{eqnarray}
  A_{lm\nu}\simeq C \int &&
  \Psi(\theta,\varphi,t)H^{(2)}(L_j\theta)\nonumber\\&&\times
  e^{-(\imath m\varphi+2\pi\nu t)}\, \theta\,\dif\theta\,\dif\varphi\,\dif t \label{inta1}\ ,\\
  B_{lm\nu}\simeq C \int&&
  \Psi(\theta,\varphi,t) H^{(1)}(L_j\theta)\nonumber\\&&\times
  e^{-(\imath m\varphi+2\pi\nu t)}\, \theta\,\dif\theta\,\dif\varphi\,\dif t\label{intb1}\
  ,
\end{eqnarray}
where $C$ is a normalisation which is given approximately by $L/(4T\Theta)$ with $\Theta$ the
range in latitude over which the integration is carried out, and $T$ the duration of the observation.

Summarized, the signal is decomposed by a spatial Han\-kel
trans\-for\-ma\-tion and a temporal Fourier transformation. The result
of this decomposition is a set of frequency spectra
$|A_{lm\nu}|^2$ and $|B_{lm\nu}|^2$ depending on
the harmonic degree $l$. Peaks occurring in these spectra are to be
identified as modes of different radial order $n$ with the
frequencies $\nu_{nl}^{\mbox{\tiny pol}}$ and $\nu_{nl}^{\mbox{\tiny
eq}}$ respectively.

Potentially differences between the mode frequencies for poleward and
equatorward propagating waves with same indices emerge due to advection. The frequency differences carry
information about the influence of the horizontal component of the
meridional circulation on the p-modes. Thus an exploration of the set
of spectra leads to a set of frequency shifts $\Delta \nu
_{nl}:=\nu_{nl}^{\mbox{\tiny pol}}- \nu_{nl}^{\mbox{\tiny eq}}$
assigned to a respective set of unique indices.

\subsection{Effect of meridional circulation on the p-modes}

In the following we provide a formula that describes the effect of a horizontal flow $\mathbf{U}$ on the p-mode frequencies.
For detailed steps to obtain this formula see~\cite{fanbraun} and references
therein.

Under the influence of a horizontal flow $\mathbf{U}$ the net frequency shift between waves
propagating poleward and equatorward is given by
\begin{equation}\label{freqdiff}
  \Delta \nu _{nl}  = \frac{l \int \limits_0^{R_{\odot}}
 (\langle{U}\rangle_{\theta}/r) K_{nl}(r) ~\dif r }{\pi \int \limits_0^{R_{\odot}} K_{nl}(r) ~ \dif r }\ ,
\label{infl}
\end{equation}
where $K_{nl}(r)$ is a depth- and mode-dependent
weighting function and
\begin{equation}
  \langle {U} \rangle _{\theta} : = \frac{1}{\theta_{\mbox{\tiny
  max}}-\theta_{\mbox{\tiny min}}} \int \limits_{\theta_{\mbox{\tiny
  min}}}^{\theta_{\mbox{\tiny max}}} {\mathbf{U}}\cdot\mbox{\boldmath$\hat{\theta}$} ~\dif \theta\ .
\end{equation}
is the flow average over the spanned polar angle.
Following~\cite{fanbraun} the frequency shift of a zonal mode due to advection
is proportional to the horizontal wavenumber and the quantity
\begin{equation} \label{u'}
  \langle U \rangle:=\Delta \nu_{ln} \pi R_{\odot}/l\ ,
  \label{eq}
\end{equation}
gives the weighted depth average of the angular mean meridional flow $\langle U\rangle_\theta/r$ multiplied by the solar radius.

Together with
\begin{equation}
\bar{\nu}:=\frac{1}{2}\left( \nu_{nl}^{\mbox{\tiny pol}} + \nu_{nl}^{\mbox{\tiny
eq}}  \right)\ ,
\end{equation}
all these considerations yield a data set
$\{\left(\langle U\rangle\!,\bar{\nu}/L\right)_i\}$, where each $i$ corresponds to a
given p-mode.

In the last step the turning points of the respective
modes are  determined by using the
relation
\begin{equation}\label{turning2}
  \frac{c^2(r_t)}{r_t^2}=\frac{(2\pi\bar{\nu})^2}{l(l+1)}\ ,
\end{equation}
with $r_t$ the depth of the inner turning point. From this we obtain the mode
penetration depth $\epsilon:=R_{\odot}-r_t$ below the
solar surface.

We note that the estimation of the integral carried out is valid only in very
shallow layers as it required approximating $\langle{U}\rangle_{\theta}/r$
by $\langle{U}\rangle_{\theta}/R_\odot$. The velocity amplitude inferred
later is therefore only reliable within the outer layers of the Sun. In
deeper layers, however, this estimation might only provide information about
the quality of the velocity profile, i.e. the sign of the weighted velocity
average.

\subsection{Data analysis}

The analysed data consist of a time series of full disk Dopplergrams recorded
by SOHO/MDI. The timeseries lasts from April 1--30, 1999 and consists
of 43200 consecutive single Dopplergrams. Each Dopplergram is given
on a 1024$\times$1024 pixel grid. The duty cycle of the time series is better than
99\%.
The data used by~\cite{fanbraun} were timeseries of MDI and GONG Dopplergrams from 1997.

Each Dopplergram is projected onto an equidistant
$\theta$-$\varphi$-lattice, referring to the standard three-dimensional
spherical coordinate system, $\theta$-centered at the solar north
pole.
After the transformation the Dopplergram is
asymmetric with respect to the equator, because at the observing time the
northern hemisphere of the Sun was better visible.
The oscillation signal is therefore better measurable on the northern
hemisphere.

Two areas on the solar surface are investigated, one in the northern the
other in the southern hemisphere. The two investigated zones are placed
symmetrically to the equator with a polar interval of $\theta_{\mbox{\tiny
tot}} = \pi/4$ and an azimuthal interval of $\varphi{\mbox{\tiny tot}} =
\pi/2$. The selected areas are centered at $\{\theta= \pi/4, \varphi=\pi/2\}$
and $\{\theta =3/4~\pi, \varphi=\pi/2\}$ respectively. The sizes of the
fields were chosen such to yield the best oscillation signal-to-noise ratio.
Moreover this allows the resolution of oscillations with low harmonic degree
$l$ in order to reach to greater depths in the solar interior. This choice is
comparable with the investigation carried out by~\cite{fanbraun}.

The geometry of the chosen fields restricts the possible resolution of the
harmonic degree to $l\ge 4$ as a lower boundary. The resolution of the
Dopplergrams yields an upper boundary of $l\le 1024$. As we assume a
$\varphi$-independent meridional circulation, only modes with $m=0$ are
investigated.

\section{Results}

\subsection{General results for both hemispheres}

\begin{figure}[htbp]
  \centering
  \includegraphics[clip,viewport=20 0 550 330, width=\columnwidth]{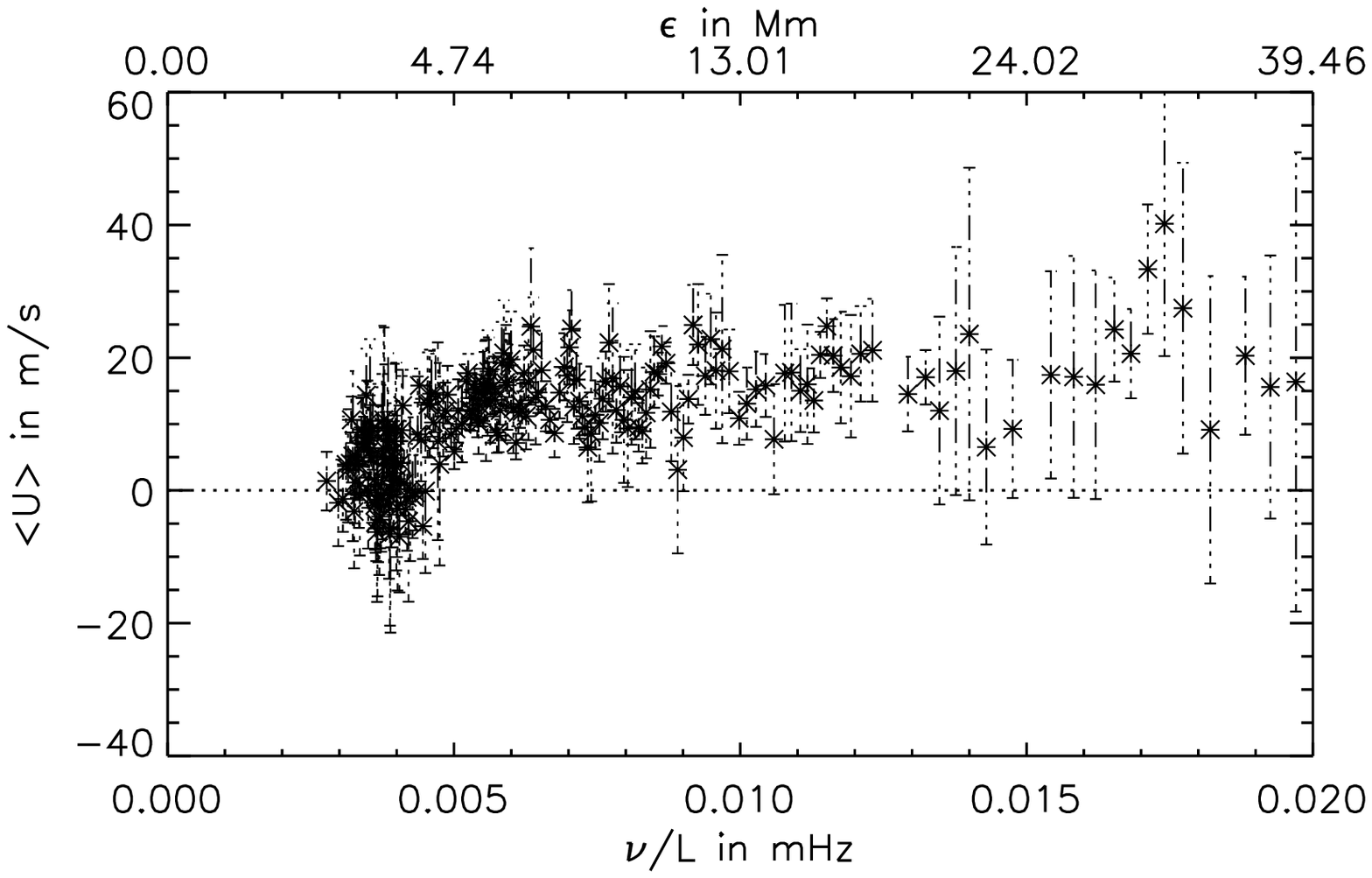}
  \includegraphics[clip,viewport=20 0 550 330, width=\columnwidth]{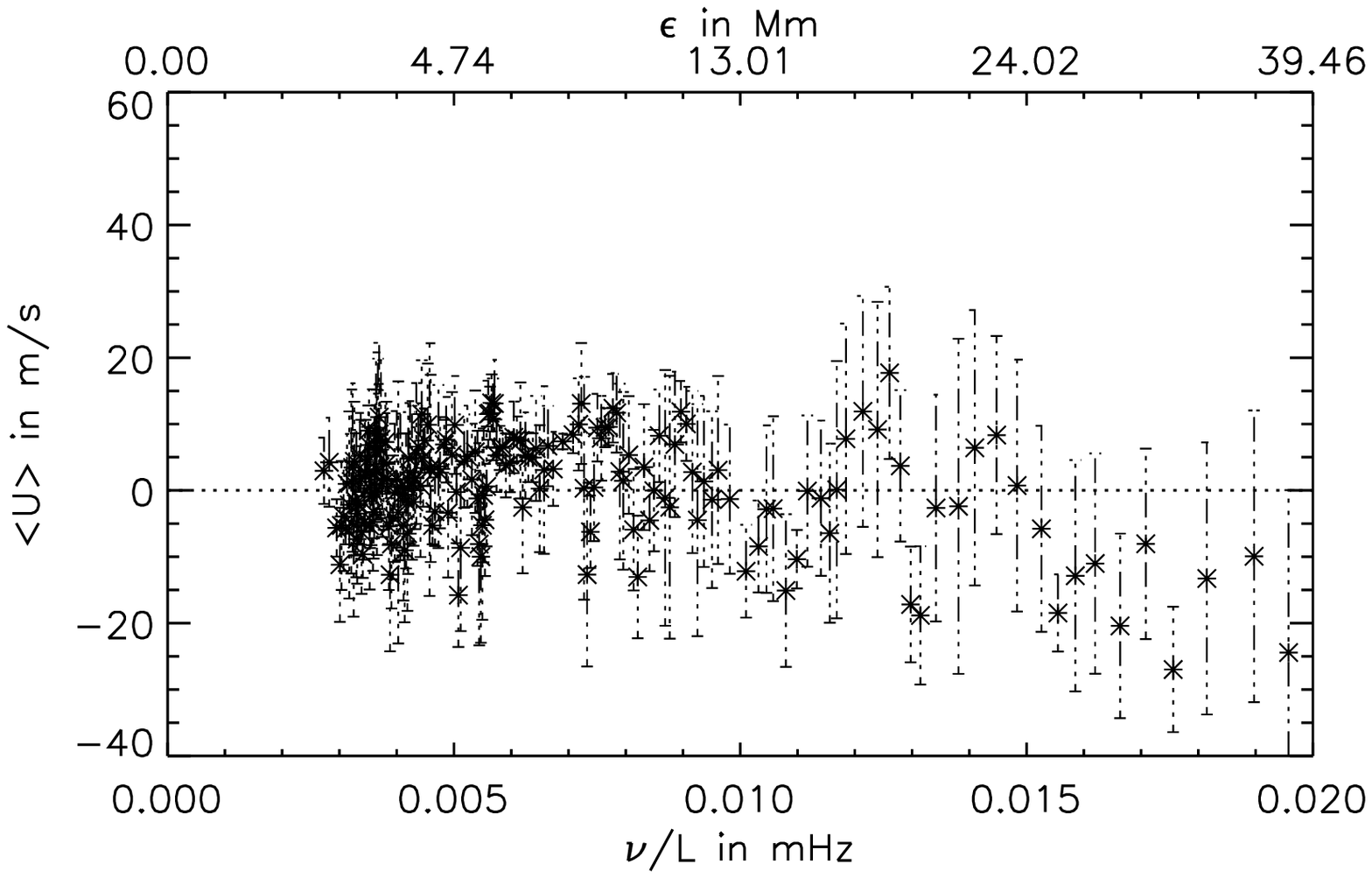}
  \caption{Estimated averaged horizontal meridional velocity on the
  northern (top) and southern (bottom) hemisphere as a function of $\nu/L$
  and as function of mode penetration depth $\epsilon$.} \label{messung}
\end{figure}

The result of our investigation are measurements of the average horizontal meridional velocity $\langle U\rangle$ as a
function of $\nu/L$. Besides the relation $\langle U \rangle \sim \nu/L$ we are interested in the
average horizontal velocity as a function of the penetration depth $\epsilon$.
Figure~\ref{messung} displays the general results.

Because of the large scatter in the results we applied a binning of the resulting
data points. In this particular case, a binning window of five points has been
chosen. This choice reflects a trade-off between a desired rather smooth profile on the
one hand and the resolution, which on the other hand should not be too coarse. The
reported error bars given in the plots below are due to the statistical scatter of the
points within one binning-window. In the plots a positive value of the velocity is
equivalent to a poleward flow on the respective hemisphere.

In particular, Fig.~\ref{messung} (top) gives the averaged horizontal meridional flow on the
northern hemisphere.
Obviously, various systematic trends are noticeable.
First, for modes penetrating between 1--4~Mm the results are distributed around $0$ m/s. This
feature has not been seen in other investigations of the meridional flow
and may
correspond to effects of granulation and supergranulation. The true origin of
this is, however, not clear.
Second, for modes probing deeper below
the solar surface there is a clear positive-valued
trend of $(15\pm 7)$~m/s. The error bars in this region are relatively
small.
Third, the observed variation in the curve might be an artefact due to
different sensitivities of the p-modes to the flow.
For the given mode penetration depths, the results agree well with meridional flow measurements by other local
helioseismology techniques like ring-diagram
analysis~(\cite{haber, zaatri}) or time-distance
helioseismology~(\cite{zhao}).

Figure~\ref{messung} (bottom) displays the results for the southern
hemisphere. As in the former case, for modes turning close to the solar surface the
points are distributed around $0$ m/s.  For modes turning further down in the deep
interior a similar general trend of the curve is observable as on the
northern hemisphere.  However, the average velocity is about $0$ m/s,
in contrast to the northern hemisphere where a value of ca. $(15\pm
7)$ m/s is obtained. Additionally the curve shows a far larger amount
of scatter, not only expressed in terms of larger error-bars but also
in the shape of the curve itself.  One reason for this difference
between the northern and southern hemisphere might be an error in
the determination of the direction of the rotation axis of the
Sun~(\cite{giles}). Another reason might arise by leakage effects from
modes with $m\not=0$ whose frequencies are shifted by differential
rotation.  In the following steps we further analyse the
measurements for the northern hemisphere data only.

We compare our results obtained for the northern hemisphere with the results
presented by Braun \& Fan (1998). We analyse a data
set set from 1999 whereas Braun \& Fan (1998) analysed a data set from 1997.
Both results agree well. Differences can only be found on small scales. We
considered only modes with an azimuthal order of $m=0$, whereas the comparable
work used a bigger set with $m \neq 0$ and applied an additional averaging.
Therefore our errorbars are bigger. Overall, this gives confidence in our
newly developed data analysis technique and the resulting measurements.

\subsection{Possible extension to greater depths}

\begin{figure}[htbp]
  \centering
  \includegraphics[clip,viewport=20 0 550 330, width=\columnwidth]{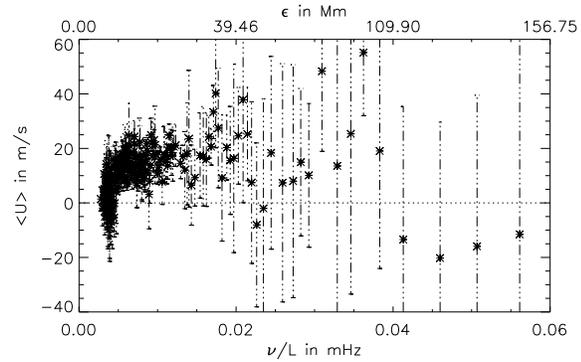}
  \caption{Estimated integrated horizontal meridional velocity on the northern
  hemisphere of the Sun as a function of $\nu/L$ and mode penetration depth $\epsilon$.} \label{tiefnord}
\end{figure}

We are able to measure frequency shifts for modes with harmonic degrees of
$l=73$ -- $1013$. The modes with the lower harmonic degree penetrate to
depths of about 150 Mm. Therefore this method might allow to derive
information about the meridional flow throughout 75\% of the convection zone.
Fig.~\ref{tiefnord} gives the averaged horizontal meridional velocity
$\langle U\rangle $ estimated by this method for modes penetrating to such depths.
Between the values of $\epsilon=20$~Mm and
$\epsilon=100$~Mm there is no clear behaviour visible. Below $\epsilon
\approx 110$~Mm the curve shows a trend to become negative-valued. However
there are only a few measurements at great depths. Moreover the results are
largely scattered which leads to the respective large error-bars.
Nevertheless, the measured frequency shifts contain information on the
meridional flow down to approximately 150 Mm.
\FloatBarrier

Based on our results we are able to suggest improvements for this technique
which we discuss in the following section.

\section{Discussion and Future Work}

We applied a Fourier-Hankel decomposition to MDI Dopplergrams as described
by~\cite{fanbraun} to estimate the average effect of the meridional
flow on solar p-modes. We find an average poleward meridional flow
velocity on the northern hemisphere in the order of 15 m/s which affects
modes penetrating in a depth range of 4--20 Mm. Occurring advection
effects on modes penetrating to shallow layers ($\epsilon \approx$ 1--4~Mm)
are interpreted as the effect of (super-) granulation. However the latter feature
was not found in other helioseismic investigations; its true origin needs to
be clarified. Furthermore frequency shifts for modes with an inner turning
point at about 150~Mm were measured.

It is encouraging that for modes turning in the upper 20~Mm of the
convection zone the inferred average meridional flow agrees well with other
independent measurements of the meridional flow velocity~(\cite{fanbraun,
giles, haber, zhao, zaatri}). Furthermore, as there is an almost perfect
agreement between our results and the results of~\cite{fanbraun} we are
sure that we were able to rebuilt their original concept.

The clear differences between the northern and southern he\-mi\-spheres
could be either due to unresolved asymmetries in the
observational setup at the given observing time or due to other
systematic errors. At least for MDI data an offset of typically about
$10$ m/s in measurements of the meridional flow between the two
hemispheres occur, independent from the method of investigation (see
e.g.~\cite{fanbraun} or \cite{zaatri}).

The presented method is only a rough
first-step-estimation of the profile of the meridional circulation.
Future investigations using the decomposition of the wavefield in
space and time and aiming at higher accuracy in the result should take
into account the spherical symmetry of the problem. Using
Legendre functions instead of Hankel functions will be a first but vital
step. Additionally the effect of leakage, potentially not only based
on a finite set of degrees (c.f.~\cite{gizon05}) but also on the
neglected influence of modes with $m \neq 0$, has to be analysed.

We conclude that the presented method -- as is -- provides a
tool for an easy-to-handle analysis of the average solar
meridional flow profile in shallow layers. Using longer time series, this
method
might allow a better frequency resolution to obtain better estimates
for the frequency shifts. Moreover, there is some freedom in choosing
the investigated range over polar angle. This might allow to assess
estimates for the average meridional flow as a function of latitude, too.

The used assumptions break down for modes penetrating to layers below 20~Mm.
Moreover the scatter in the data is high at great penetration depths.
Nevertheless, we find a negative-value trend in the estimation of the average
meridional flow for p-modes probing deeper than 110~Mm. Such a negative value
might give a hint to an equatorwards directed meridional flow in the layers
above. However, before concluding on the meridional return flow all
entrapments in the data analysis need to be resolved.

Most important, a proper inversion of the measured frequency shifts needs to
be carried out to estimate the velocity profile over a certain depth range.
This will require to deal with the forward problem properly. A normal mode
approach needs to be carried out to obtain not only a proper estimation of
the effect of all three spatial components of the meridional flow on the mode
frequencies but also to obtain integral kernels. Then it will be interesting
to see, whether enough data and information could be collected to construct
averaging kernels at greater depths. If all these suggested improvements
are incorporated, then
the meridional flow could be studied in better detail as function of depth,
latitude and time with this method.

\acknowledgements

This work was initiated during the ISSI (International Space Science
Institute, Bern, Switzerland) workshop ''Observations and Models of the Solar Cycle''
in March 2005. The authors are very grateful to A.G. Kosovichev for helpful
comments on a draft of this paper. The authors thank D. Braun for helpful
discussions. Lars Krieger acknowledges support from HELAS to visit the
Max-Planck-Institut f\"ur Sonnensystemforschung for collaboration on this
project. Participation of Lars Krieger and Markus Roth at the workshop in Nice was
supported by HELAS. The European Helio- and Asteroseismology Network (HELAS)
is funded by the European Union's Sixth Framework Programme.


\newpage
\begin{thebibliography}{}

\bibitem[Bogdan et al. (1998)]{bogdan} Bogdan, T.J., Braun, D.C., Lites, B.W., Thomas,
  J.H.: 1998, ApJ 492, 379

\bibitem[Brandenburg, Moss \& Tuominen (1992)]{bb}Brandenburg, A., Moss, D. Tuominen,
  I.: 1992, A\&A 265, 328

\bibitem[Braun (1995)]{braun95} Braun, D.C.: 1995, ApJ 451, 895

\bibitem[Braun, Duvall, \& Labonte (1987)]{braun87} Braun, D.C., Duvall Jr., T.L., Labonte,
  B.J.: 1987, ApJ 319, L27

\bibitem[Braun, Duvall \& Labonte (1988)]{braun88}Braun, D.C., Duvall Jr., T.L., Labonte,
  B.J.: 1988, ApJ 335, 1015

\bibitem[Braun et al. (1992)]{braun92} Braun, D.C., Duvall Jr.,T.L., Labonte,
  B.J., Jefferies, S.M., Harvey, J.W., Pomerantz, M.A.: 1992, ApJ 391, L113


\bibitem[Braun \& Fan (1998)]{fanbraun}Braun, D.C., Fan, Y.: 1998, ApJ 508, L105

\bibitem[Choudhuri, Sch\"ussler \& Dikpati (1995)]{choud}Choudhuri, A.R.; Sch\"ussler, M.;
  Dikpati, M.: 1995, A\& A 303, L29

\bibitem[Dikpati \& Charbonneau (1999)]{dik2}Dikpati, M., Charbonneau, P.: 1999, ApJ 518, 508

\bibitem[Dikpati \& Gilman (2006)]{dik1}Dikpati, M., Gilman, P.A.: 2006, ApJ 649, 498

\bibitem[Giles et al. (1997)]{giles97} Giles, P.M., Duvall, T.L.,
Jr., Scherrer, P.H., Bogart, R.S.: 1997, Nature, 390, 52

\bibitem[Giles (2000)]{giles}Giles, P.M.: 2000, PhD Thesis

\bibitem[Gizon \& Birch (2005)]{gizon05} Gizon, L., Birch, A.C.: 2005, Living Rev. Sol.
Phys., 2, 6

\bibitem[Haber et al. (2002)]{haber} Haber, D., Hindman, B.W., Toomre, J., Bogart, R.S.,
  Larsen, R.M., Hill, F.: 2002, ApJ 570, 855

\bibitem[Hathaway (1996)]{hatha}Hathaway, D.H.: 1996, ApJ 460, 1027

\bibitem[Komm, Howard \& Harvey (1993)]{komm}Komm, R.W., Howard, R.F., Harvey, J.W.:
  1993, Sol. Phys. 147, 207

\bibitem[Latushko (1996)]{latu}Latushko, S.: 1996, Sol. Phys. 163, 241

\bibitem[Snodgrass \& Daily (1996)]{snod}Snodgrass, H.B., Daily, S.B.: 1996, Sol. Phys. 163, 21

\bibitem[Zaatri et al. (2006)]{zaatri} Zaatri, A., Komm, R., Gonzalez Hernandez, I.,
  Howe, R., Corbard, T.: 2006, Sol. Phys. 236, 227

\bibitem[Zhao \& Kosovichev (2004)]{zhao} Zhao, J., Kosovichev, A.G.: 2004, ApJ 603, 776

\end{thebibliography}
\end{document}